\documentclass[runningheads]{llncs}
\usepackage{graphicx}
\usepackage{hyperref}
\usepackage{amsmath}
\newcommand{\now}{\textit{present point}} 
\newcommand{\shortname}{LoMaR} 
\newcommand{\sectf}{visit} 

\begin{document}
\title{Longitudinal Mammogram Risk Prediction}
\titlerunning{\shortname}
\author{Batuhan K. Karaman, MS\inst{1,2} \and
Katerina Dodelzon, MD\inst{2} \and
Gozde B. Akar, PhD\inst{3} \and
Mert R. Sabuncu, PhD\inst{1,2}}
\authorrunning{Karaman et al.}
\institute{Cornell University and Cornell Tech, New York, NY 10044, USA \and
Weill Cornell Medicine, New York, NY 10021, USA \and
Middle East Technical University, Ankara 06800, Turkey \\
}
\maketitle              
\begin{abstract}
Breast cancer is one of the leading causes of mortality among women worldwide. 
Early detection and risk assessment play a crucial role in improving survival rates. 
Therefore, annual or biennial mammograms are often recommended for screening in high-risk groups.
Mammograms are typically interpreted by expert radiologists based on the Breast Imaging Reporting and Data System (BI-RADS), which provides a uniform way to describe findings and categorizes them to indicate the level of concern for breast cancer. 
Recently, machine learning (ML) and computational approaches have been developed to automate and improve the interpretation of mammograms.
However, both BI-RADS and the ML-based methods focus on the analysis of data from the present and sometimes the most recent prior visit.
While it is clear that temporal changes in image features of the longitudinal scans should carry value for quantifying breast cancer risk, no prior work has conducted a systematic study of this.
In this paper, we extend a state-of-the-art ML model~\cite{yala_2021_toward} to ingest an arbitrary number of longitudinal mammograms and predict future breast cancer risk. 
On a large-scale dataset, we demonstrate that our model, {\shortname}, achieves state-of-the-art performance when presented with only the present mammogram.
Furthermore, we use {\shortname} to characterize the predictive value of prior visits. 
Our results show that longer histories (e.g., up to four prior annual mammograms) can significantly boost the accuracy of predicting future breast cancer risk, particularly beyond the short-term. 
Our code and model weights are available at \href{https://github.com/batuhankmkaraman/LoMaR}{https://github.com/batuhankmkaraman/{\shortname}}.

\keywords{Breast Cancer Risk Prediction \and Longitudinal Data \and Transformer Neural Networks}
\end{abstract}
\section{Introduction}
Breast cancer is the most prevalent cancer worldwide~\cite{ref_1}, making population-level screening essential for early detection and improved treatment outcomes. 
Mammography (low-dose X-Ray imaging of the breasts) stands as the principal method among various breast cancer screening techniques, including MRI and ultrasound, due to its accessibility, effectiveness, and reliability. 
Accurate risk prediction at the time of mammography is essential for adjusting screening intervals to suit individual patients and for determining the need for supplementary modalities for comprehensive evaluation.
Traditional models for breast cancer risk assessment, such as the Tyrer-Cuzick (TC) and the Breast Cancer Surveillance Consortium (BCSC), utilize specific risk factors, including mammogram-derived expert-defined features such as 
breast density information, to estimate a patient's risk~\cite{ref_21}. 
Yet, 
the reliability and utility of these predictions can vary.
In response, recent advances have seen the emergence of machine learning (ML) based algorithms that analyze mammographic data for breast cancer risk prediction. 
These approaches have shown promising results, often surpassing conventional risk models in their ability to predict both immediate and long-term risk of breast cancer~\cite{ref_2,ref_3,ref_5,ref_6,ref_9,ref_12,ref_17,ref_despina}.


However, the full potential of longitudinal data is often not realized in contemporary approaches. 
Most studies, including those using advanced deep learning techniques, predominantly analyze only the present mammogram, possibly together with 
limited prior visit, failing to take advantage of the comprehensive longitudinal mammographic history available. 
Despite this limitation, these models still achieve state-of-the-art prediction accuracy~\cite{ref_9},\cite{ref_17}. 
Recently, \cite{ref_6} posited that a more extensive analysis that includes longer periods of screening data could lead to a richer understanding of the patterns and progression of breast cancer. 
Our paper aims to provide a systematic analysis of the utility of a broader range of historical data in quantifying breast cancer risk.


In this work, we present a Longitudinal Mammogram Risk model ({\shortname}), which builds on~\cite{yala_2021_toward} and implements a transformer architecture~\cite{vaswani_2017_attention,dosovitskiy2020image}, coupled with a convolutional feature extractor~\cite{he_2016_deep}, to ingest 2-view mammograms collected from an arbitrary number of (present and past) visits and predict breast cancer risk.
We use a comprehensive dataset to evaluate our model and achieve state-of-the-art prediction performance, even with just the latest mammogram.
Additionally, we explore how {\shortname} benefits from the information provided by previous visits. 
Our analysis highlights the critical importance of longitudinal data in clinical environments, illustrating its critical role in enhancing the early detection of breast cancer.

\section{Materials and Methods}


\subsection{Dataset}
All participants in this study are from the Karolinska case-control (CSAW-CC) dataset~\cite{dataeuropaeu}, derived from the Cohort of Screen-Aged Women~\cite{dembrower_2019_a} in order to perform AI research to improve screening, diagnostics and prognostics of breast cancer.
Karolinska dataset consists of women aged 40 to 74 who underwent mammography screenings over multiple years between 2008 and 2016. 
Each mammogram contains four X-ray scans from the craniocaudal (CC) and mediolateral oblique (MLO) views for both the left and right breasts, with all images captured using Hologic machines. 
The dataset includes a total of 19,328 mammograms, with 1,413 resulting in a cancer diagnosis, from 7,353 patients.
Distribution of cancer-related labels and our image preprocessing steps are included in Sections S.1 and S.2 of the supplementary material, respectively.

\subsection{Model} 
\begin{figure}[htb]
\includegraphics[width=\textwidth]{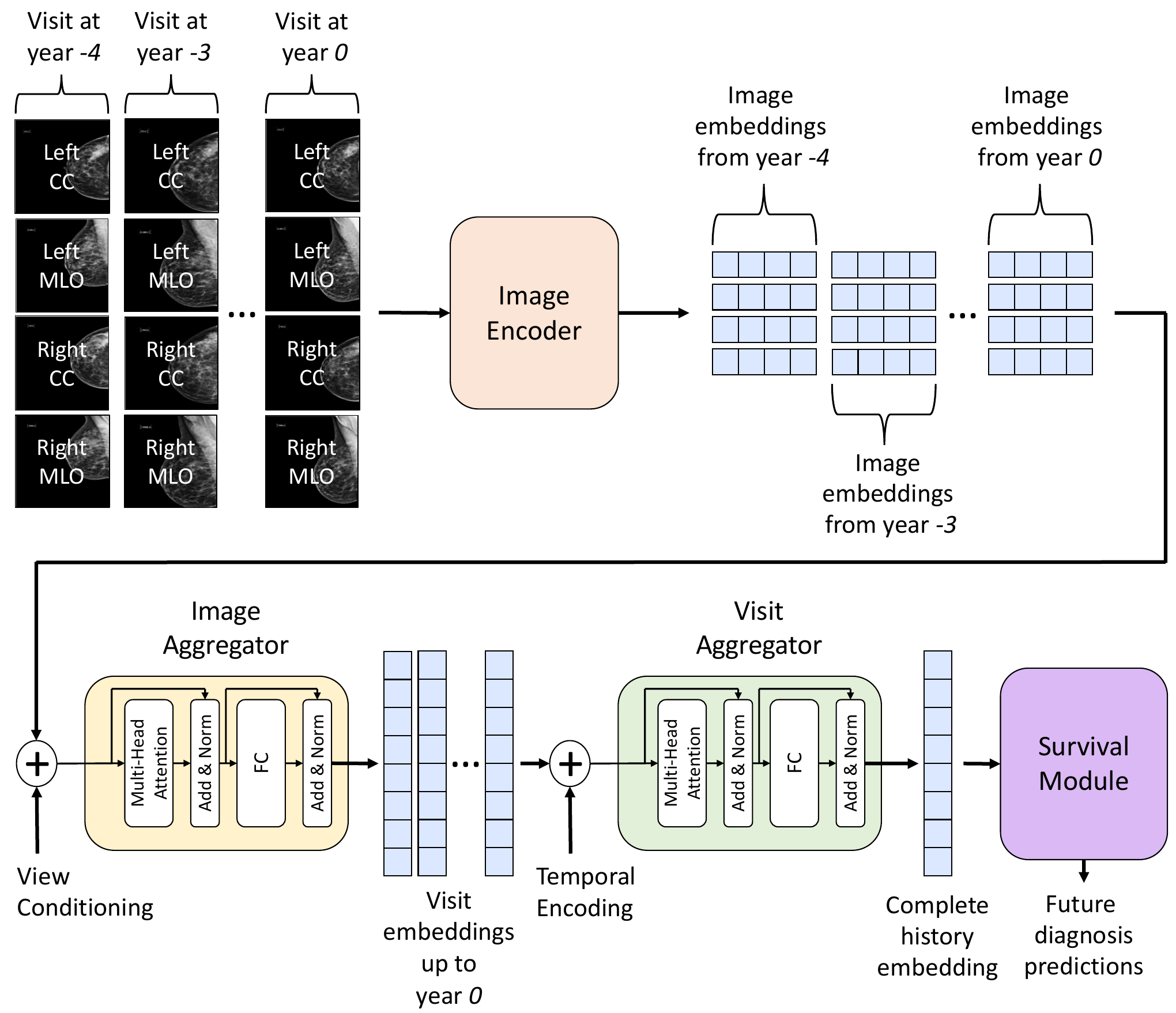}
\caption{Schematic representation of our Longitudinal Mammogram Risk model ({\shortname}), where year 0 represents the current point in time.} 
\label{figmodel}
\end{figure}

We focus on predicting an individual's future breast cancer diagnosis status using mammograms obtained from present and prior visits. 
To effectively harness the longitudinal sequence of visits and make future predictions for consecutive follow-up years after the present mammogram, we propose a transformer-based neural network which we refer to as {\shortname}, which stands for Longitudinal Mammogram Risk model.
{\shortname} extends the state-of-the-art Mirai model that only ingests the present mammogram~\cite{yala_2021_toward}. 
The detailed schematic of {\shortname} is depicted in Figure~\ref{figmodel}.

{\shortname} begins by processing each mammogram, comprising four 2D X-ray images of both left and right breasts in craniocaudal (CC) and mediolateral oblique (MLO) views, through a shared image encoder to produce distinct embeddings for each image. 
The image aggregator, a transformer encoder~\cite{dosovitskiy2020image}, then fuses the embeddings from the same visit, creating a visit embedding that integrates the information across the different views and both sides. 
These visit embeddings are then compiled by the {\sectf} aggregator, a subsequent transformer that processes each visit embedding as an individual token and captures the chronological progression of the visits.
This results in a single, extensive embedding reflecting the entire longitudinal history. 
Finally, the survival module takes this embedding to predict the future cancer risk of the patient, employing a strategy that accounts for the progression of the disease over time.

In terms of network architecture, we utilize a ResNet-based convolutional neural network (CNN)~\cite{he_2016_deep} as the image encoder, followed by two transformer encoders that facilitate the aggregation of images and visits. 
The CNN and image aggregator employed in our model are identical to those described in~\cite{yala_2021_toward}. 
More information about these two modules can be found in Sections S.3 of the supplementary material



The {\sectf} aggregator incorporates fixed sine and cosine positional encodings as temporal encodings at its input, as introduced in~\cite{vaswani_2017_attention}, which enables it to interpret the temporal ordering of visits. 
It processes the sequence through its self-attention block and employs average sequence pooling to distill the patient's entire mammographic history into a comprehensive embedding.


Our survival module is an additive hazard layer, similar to those found in~\cite{ref_17} and~\cite{yala_2021_toward}. 
It leverages the complete history embedding, denoted as $m$, for predicting future cancer risk. 
The module determines a baseline risk $B(m)$ and computes incremental risks for each subsequent year using linear layers. 
The incremental risk outputs are passed through ReLU activations. 
The cumulative $k$-year risk $P(t_{\text{cancer}} = k | m)$ is the sum of these risks, represented by the sigmoid of the sum of $B(m)$ and yearly hazards $H_i(m)$:
\begin{equation}
\label{eqsurv}
P(t_{\text{cancer}} = k | m) = \sigma(B(m) + \sum_{i=1}^{k} H_i(m)),
\end{equation}
guaranteeing risk prediction growth over time.

\subsection{Training} 
\label{sectraining}
Despite their effectiveness, transformers are data-intensive and prone to overfitting due to their self-attention mechanisms; hence, we employed a training strategy with multiple techniques to mitigate overfitting and improve model performance.
We consider every visit within the training data as a present-time reference point, denoted as {\now}. 
For each visit identified as {\now}, we gather the patient’s historical data, reaching back up to four years. 
This results in the most extensive training sequence comprising five visits: the visit four years before {\now}, three years before, two years before, one year before, and the visit coinciding with {\now}. 
We then track the subsequent diagnoses for the five following years after {\now}, creating a comprehensive 10-year window around each {\now} instance.
By constructing these 10-year progression trajectories, we substantially expand the size of our training and validation datasets. 
To prevent information leakage, we ensure that all splits are at the individual subject level.

For the image encoder and image aggregator, we utilized the pretrained encoder and aggregator from~\cite{yala_2021_toward}, with both components frozen during training to retain their pre-learned features~\cite{ref_11}.
To address the imbalance in label distribution over the five-year follow-up horizon during training the {\sectf} aggregator and survival module, we use the reweighted cross-entropy loss function from~\cite{karaman_2022_machine} by adapting it for binary classification.
The weights for each sample point are calculated based on the expanded datasets.

\subsection{Evaluation}
In evaluation, our main objective is to assess the influence of a patient's longitudinal mammographic history availability on the predictions made by our model.
Notably, {\shortname} is designed to handle incomplete histories, allowing for predictions with limited or no past data. 
We explore this by testing the model's performance across various history scenarios, simulating different durations and frequencies of patient history by selectively including or omitting visits. 
This methodology helps us understand the model's adaptability to real-world clinical settings where patient data may be incomplete or irregularly recorded.


In real-world longitudinal studies, biases in subject recruitment and follow-up are prevalent and can markedly affect the validity of the findings~\cite{yu_2020_one,tasci_2022_bias}.
To counter potential biases, we employ an inference strategy designed to mitigate these discrepancies.
We primarily use the area under the receiver operating characteristic (ROCAUC) metric, which accounts for the imbalance in the labels.
To compute the ROCAUC score for a specific follow-up year (e.g, 2 years into the future) and longitudinal history scenario (e.g., with visits from present, and 1 year ago), we start by randomly selecting a single visit to represent {\now} for each test subject. 
This selection is made from all instances associated with a diagnosis in the designated follow-up year.
The chosen instances of {\now} are then combined to form a `random' pseudo test set. 
It is important to note that within this pseudo test set, each subject contributes a single follow-up diagnosis for prediction based on a specific historical instance, ensuring that all sample points in the pseudo test set are independent.
After obtaining predictions using the visits that fit the specified longitudinal history scenario, we record the ROCAUC value. 
This process is repeated for multiple random pseudo test sets. 
Finally, we calculate the average of the ROCAUC values obtained from these pseudo test sets.
This average ROCAUC score represents the comprehensive evaluation for the specific follow-up year and longitudinal history scenario pair. 
We note that the purpose of repeating the operation for multiple pseudo test sets is to broaden the range of disease progression trajectories we use in our evaluation and thus mitigate the potential effects of the aforementioned biases on our results.

\section{Experiments}

\subsection{Implementation Details}

We split the data into training and testing sets in an 80-20 ratio using a randomized, diagnosis-stratified approach, repeated 10 times. 
The results shown are the average of these iterations with 95\% confidence intervals. 
Within each split, 25\% of the training data was used for validation to determine early stopping based on validation loss. 
We optimized {\shortname}'s architecture using grid search across the splits, selecting the best architecture based on validation set performance. 
Fixed and tunable hyperparameters are detailed in Section S.4 of the supplementary material. ROCAUC calculations were performed using 100 random pseudo test sets.

\subsection{Results}
Table~\ref{tab1} presents C-index and ROCAUC scores for various models, including our {\shortname} model tested across different longitudinal history durations and data collection frequencies, with Image-Only DL~\cite{ref_12} and Mirai~\cite{yala_2021_toward} as benchmarks. 
We calculate the C-index using the same method as we use for the ROCAUC.
Image-Only DL and Mirai use the same image encoder and visit aggregator as {\shortname}, while Mirai is considered as the state-of-the-art in the Karolinska dataset.
Our proposed model, {\shortname}, demonstrates superior performance over the Image-Only DL and Mirai models, even when no historical data are utilized, as shown by the ROCAUC scores across all follow-up years except follow-up year 2. 
For instance, with a ROCAUC score of 0.92 for the 1-year follow-up, our model significantly surpasses the Image-Only DL's score of 0.83 and Mirai's score of 0.90.

{
\begin{table}[htb]
\caption{
C-index and ROCAUC scores of {\shortname} and other existing models from the literature. 
{\shortname} is evaluated by creating various longitudinal history durations (in years) and frequency scenarios in the test set. The notation $^{\star}$ represents evaluations with annual data collection frequency, while $^{\dagger}$ denotes evaluations with biennial data collection for the corresponding history duration.
}
\label{tab1}
\fontsize{8.5}{12}\selectfont
\begin{tabular}{|c|c|c|c|c|c|c|c|}
\hline
~ & History & ~ & \multicolumn{5}{c|}{Follow-up year ROCAUC}\\\cline{4-8}
Model  & duration & C-index & 1-year & 2-year & 3-year & 4-year & 5-year\\
\hline
Image-Only & 0 & 0.75 & 0.83 & 0.79 & 0.75 & 0.73 & 0.71 \\
DL \cite{ref_12} & ~ & (0.73–0.77) & (0.81–0.86)  & (0.77–0.81) & (0.73–0.77) & (0.71–0.75) & (0.69–0.73)\\
\hline
Mirai \cite{yala_2021_toward} & 0 & 0.81 & 0.90 & 0.86 & 0.82 & 0.80 & 0.78 \\
 & ~ & (0.79–0.82) & (0.89–0.92)  & (0.84–0.88) & (0.80–0.84) & (0.79–0.82) & (0.76–0.80)\\
 \hline
{\shortname} & 0 & 0.81 & 0.92 & 0.84 & 0.84 & 0.82 & 0.81 \\
(Ours) & ~ & (0.78–0.83) & (0.90–0.94)  & (0.83–0.86) & (0.83–0.85) & (0.81–0.83) & (0.80–0.82)\\
 \hline
{\shortname} & 1$^{\star}$ & 0.82 & 0.92 & 0.85 & 0.84 & 0.82 & 0.81 \\
(Ours) & ~ & (0.78–0.83) & (0.90–0.94)  & (0.83–0.86) & (0.83–0.85) & (0.81–0.83) & (0.80–0.82)\\
 \hline
{\shortname} & 2$^{\star}$ & 0.82 & 0.92 & 0.84 & 0.84 & 0.82 & 0.83 \\
(Ours) & ~ & (0.78–0.83) & (0.90–0.94)  & (0.83–0.86) & (0.83–0.85) & (0.81–0.83) & (0.82–0.84)\\
 \hline
{\shortname} & 3$^{\star}$ & 0.82 & 0.92 & 0.84 & 0.84 & 0.83 & 0.85 \\
(Ours) & ~ & (0.78–0.83) & (0.90–0.94)  & (0.83–0.86) & (0.82–0.85) & (0.82–0.84) & (0.84–0.86)\\
 \hline
{\shortname} & 4$^{\star}$ & 0.82 & 0.92 & 0.84 & 0.84 & 0.85 & 0.86 \\
(Ours) & ~ & (0.78–0.83) & (0.90–0.93)  & (0.83–0.86) & (0.83–0.86) & (0.84–0.86) & (0.85–0.87)\\
\hline
{\shortname} & 4$^{\dagger}$ & 0.82 & 0.92 & 0.84 & 0.84 & 0.84 & 0.84\\
(Ours) & ~ & (0.78–0.83) & (0.90–0.93)  & (0.83–0.86) & (0.83–0.86) & (0.84–0.86) & (0.85–0.87)\\
\hline
\end{tabular}
\end{table}
}

\subsubsection{Impact of longitudinal mammographic history duration.}

Examining the ROCAUC scores for various longitudinal history durations in Table~\ref{tab1}, we note that {\shortname} shows a progressive improvement in long-term prediction performance as the history duration increases.
For instance, when four years of past mammogram data are included, the 5-year ROCAUC score of {\shortname} improves to 0.86, a substantial increase from the 0.81 achieved with no history. 
This improvement is consistent with the clinical understanding that a deeper historical context can provide more insight into the patient's breast health over time.
It's important to also note that the earlier years' prediction performance remains stable regardless of the history duration included.
This suggests that the initial changes are typically discernible in the mammograms at {\now}.

\subsubsection{Impact of longitudinal data collection frequency.}
{
The comparison of longitudinal history durations 4$^{\star}$ and 4$^{\dagger}$ in Table~\ref{tab1} highlights the importance of more frequent screening. 
With an annual screening frequency (4$^{\star}$), the model achieves a 5-year ROCAUC score of 0.86, whereas with a biennial frequency (4$^{\dagger}$), the score is slightly lower at 0.84. 
This suggests that more frequent data collection can enhance the model's ability to predict long-term breast cancer risk, underscoring the value of annual screenings in improving risk assessment accuracy.
}

We observe a similar pattern of results when we exclude patients who have a cancer that was confirmed within 6 months of the \now~(see Section S.5 of the supplementary material). In these results, notably, the performance boost of incorporating longitudinal history becomes even more pronounced.  

\subsubsection{Improved localization of cancer with past mammograms.}
\begin{figure}[htb]
\includegraphics[width=\textwidth]{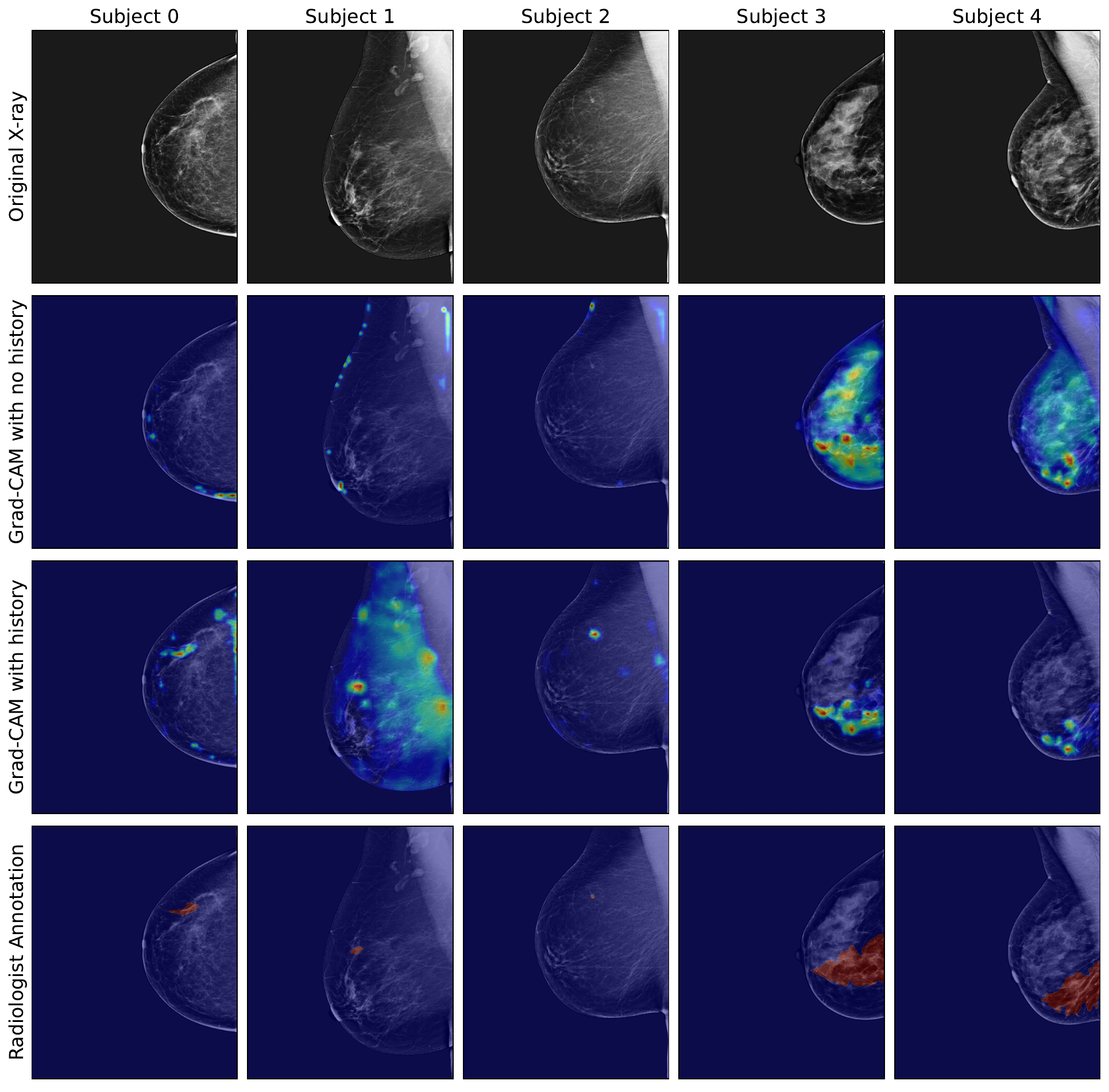}
\caption{Grad-CAM visualization of {\shortname} with mammograms from five representative test subjects with (third row) and without (second row) prior mammograms.} 
\label{figcam}
\end{figure}

{
In Fig.~\ref{figcam}, we visualize Grad-CAM saliency maps~\cite{selvaraju2017grad} for the {\now} mammograms of a representative set of test subjects diagnosed with breast cancer. These subjects are examples of cases where historical mammograms enabled {\shortname} to correct its initial prediction.
The bottom row shows the expert-defined annotations of the cancer lesion.
Comparing the second row (only based on \now) to third row (with prior visits included), we observe that the historical mammograms were critical in localizing the suspicious lesion in the scan and making the correct prediction.
}

\section{Conclusion}
In this paper, we extend a state-of-the-art ML model~\cite{yala_2021_toward} to ingest an arbitrary number of longitudinal mammograms and predict future breast cancer risk. 
We use our model {\shortname} to characterize the predictive value of prior mammograms. 
Our results show that longer histories (e.g., up to four prior annual mammograms) can significantly boost the accuracy of predicting future breast cancer risk, particularly beyond the short-term. 
This improved accuracy is due to the better localization of the suspicious lesion.

%
%
%
\bibliographystyle{splncs04}
\bibliography{mybibliography}

\newpage
\renewcommand{\thesubsection}{S.\arabic{subsection}}
\renewcommand{\thetable}{ST-\arabic{table}}
\title{Supplementary Material for Longitudinal Mammogram Risk Prediction}
\titlerunning{Supplementary Material for {\shortname}}
\author{Batuhan K. Karaman, MS\inst{1,2} \and
Katerina Dodelzon, MD\inst{2} \and
Gozde B. Akar, PhD\inst{3} \and
Mert R. Sabuncu, PhD\inst{1,2}}
\authorrunning{Karaman et al.}
\institute{Cornell University and Cornell Tech, New York, NY 10044, USA \and
Weill Cornell Medicine, New York, NY 10021, USA \and
Middle East Technical University, Ankara 06800, Turkey \\
}
\maketitle              

\subsection{Additional information about Karolinska dataset}
Refer to Table~\ref{tab3}.
{
\begin{table}[htb]
\caption{
Distribution of follow-up times and times until cancer diagnosis for examinations in the Karolinska dataset.
}
\label{tab3}

\begin{tabular}{|p{1.1cm}|p{1.1cm}|p{1.1cm}|p{1.1cm}|p{1.1cm}|p{1.1cm}|p{1.1cm}|p{1.1cm}|p{1.1cm}|p{1.1cm}|}
\hline
\multicolumn{5}{|l|}{Number of exams with minimum $n$ years} & \multicolumn{5}{l|}{Number of exams followed by a cancer } \\
\multicolumn{5}{|l|}{of screening followup} & \multicolumn{5}{l|}{diagnosis within $n$ years} \\
\hline
$n$=1 & $n$=2 & $n$=3 & $n$=4 & $n$=5 & $n$=1 & $n$=2 & $n$=3 & $n$=4 & $n$=5 \\
\hline
19328 & 16148 & 12873 & 9578 & 6530 & 517 & 681 & 1040 & 1181 & 1413 \\
\hline
\end{tabular}


\end{table}
}

\subsection{Image Processing}
To ready images for the encoder, we resize them to 1664 by 2048 pixels and align them to the left for uniformity. 
We then normalize them with mean and standard deviation values defined for the image encoder, consistent across training, validation, and test sets. 
Lastly, we convert single-channel images to pseudo-RGB by replicating them across three channels.

\subsection{Details of Image Encoder and Image Aggregator}
The CNN and image aggregator employed in our model are identical to those described in Mirai, with the CNN being a ResNet-18 followed by a global max pooling layer to produce 1D image embeddings. 
At its input, the image aggregator enhances the embeddings by conditioning them on their specific views (CC/MLO) and laterality (left/right) using learned non-parameterized positional embeddings. 
An affine transformation is applied to each image embedding $\mathbf{x}$ as follows:
\begin{equation}
    h = (W_{\text{scale}}\mathbf{e}) \odot \mathbf{x} + (W_{\text{shift}} \mathbf{e}),
\label{eqcond}
\end{equation}
where $\odot$ is the dot-product, and $\mathbf{e}$ represents the unique 1D positional embedding for each view and laterality, resulting in four distinct instances. 
The matrices $W_{\text{scale}}$ and $W_{\text{shift}}$ are two-dimensional and fixed across all views and lateralities, ensuring uniform scaling and shifting of the embeddings. 
Following this conditioning, the image aggregator uses a self-attention block to process the embeddings and uses attention pooling implemented with a linear layer followed by softmax activation to produce a singular visit representation.

\subsection{Additional Experimental Details}

During training, we used Adam with a learning rate of \(1e-3\) and implemented dropout with a probability of $0.25$ at three points: before, within, and after the {\sectf} aggregator.  
The aggregator itself consists of a single self-attention block. 
For architectural fine-tuning and learning-related hyperparameter adjustments, we conducted a grid search over three key aspects: the embedding dimension and the number of attention heads within the aggregator, and the L2 regularization applied to the model's weights and biases. 
The values we explored were \{128, 256, 512\} for the embedding dimension, \{1, 4, 8\} for the number of heads, and \{1e-4, 1e-5, 1e-6\} for the L2 rate.

\subsection{Additional Results}
Refer to Table~\ref{tab2}.
{
\begin{table}[htbp]
\caption{
C-index and ROCAUC scores for {\shortname} and other existing models, excluding cancers confirmed within 6 months post-screening. 
Symbols $^{\star}$ and $^{\dagger}$ denote evaluations with annual and biennial longitudinal data frequencies, respectively.
}
\label{tab2}
\begin{tabular}{|c|c|c|c|c|c|c|c|}
\hline
~ & History & ~ & \multicolumn{4}{c|}{Follow-up year ROCAUC}\\\cline{4-7}
Model  & duration & C-index & 2-year & 3-year & 4-year & 5-year\\
\hline
Image-Only & 0 & 0.67 & 0.66 & 0.68 & 0.66 & 0.64\\
DL & ~ & ~ & ~  & ~ & ~ & ~\\
\hline
Mirai & 0 & 0.71 & 0.72 & 0.73 & 0.73 & 0.71\\
 & ~ & ~ & ~  & ~ & ~ & ~\\
 \hline
{\shortname} & 0 & 0.70 & 0.73 & 0.73 & 0.73 & 0.72\\
(Ours) & ~ & ~ & ~  & ~ & ~ & ~\\
 \hline
{\shortname} & 1$^{\star}$ & 0.71 & 0.73 & 0.74 & 0.73 & 0.72\\
(Ours) & ~ & ~ & ~  & ~ & ~ & ~\\
 \hline
{\shortname} & 2$^{\star}$ & 0.71 & 0.73 & 0.73 & 0.73 & 0.75\\
(Ours) & ~ & ~ & ~  & ~ & ~ & ~\\
 \hline
{\shortname} & 3$^{\star}$ & 0.72 & 0.73 & 0.73 & 0.75 & 0.79\\
(Ours) & ~ & ~ & ~  & ~ & ~ & ~ \\
 \hline
{\shortname} & 4$^{\star}$ & 0.74 & 0.74 & 0.75 & 0.79 & 0.81\\
(Ours) & ~ & ~ & ~  & ~ & ~ & ~ \\
\hline
{\shortname} & 2$^{\dagger}$ & 0.71 & 0.73 & 0.73 & 0.73 & 0.75 \\
(Ours) & ~ & ~ & ~  & ~ & ~ & ~ \\
\hline
{\shortname} & 4$^{\dagger}$ & 0.73 & 0.73 & 0.75 & 0.76 & 0.78 \\
(Ours) & ~ & ~ & ~  & ~ & ~ & ~ \\
\hline
\end{tabular}
\end{table}
}

\end{document}